\documentclass[nofootinbib,preprint,preprintnumbers,a4paper,11pt]{revtex4}
\usepackage{epsfig,footnote}
\usepackage{ulem}
\usepackage{color}
\usepackage{array}
\usepackage{amssymb}
\usepackage{amsmath}
\usepackage{graphicx,subfigure}
\usepackage{longtable}
\usepackage{verbatim}
\usepackage{amsfonts}
\usepackage{hyperref}
\usepackage{multirow}

\begin{document}

\title{Sum rules for $CP$ asymmetries of charmed baryon decays \\ in the $SU(3)_F$ limit}

\author{Di Wang}\email{dwang15@lzu.edu.cn}
\affiliation{%
School of Nuclear Science and Technology,  Lanzhou University,
Lanzhou 730000, People's Republic of China}

\begin{abstract}

Motivated by the recent LHCb observation of $CP$ violation in charm,
we study $CP$ violation in the charmed baryon decays. A simple method to search for the $CP$ violation relations in the flavor $SU(3)$ limit, which is associated with a complete interchange of $d$ and $s$ quarks, is proposed. With this method, hundreds of $CP$ violation sum rules in the doubly and singly charmed baryon decays can be found. As examples, the $CP$ violation sum rules in two-body charmed baryon decays are presented. Some of the $CP$ violation sum rules could help the experiment to find better observables. As byproducts, the branching fraction of $\Xi^+_c\to pK^-\pi^+$ is predicted to be $(1.7\pm 0.5)\%$ in the $U$-spin limit and the fragmentation-fraction ratio is determined to be $f_{\Xi_b}/f_{\Lambda_b}=0.065\pm 0.020$ using the LHCb data.
\end{abstract}

\maketitle

%{\small{\tableofcontents}}
\section{Introduction}

Very recently, the LHCb Collaboration observed the $CP$ violation in the charm
sector \cite{Aaij:2019kcg},  with the value of
\begin{align}
\Delta A_{CP} \equiv A_{CP}(D^0\to K^+K^-)  - A_{CP}(D^0\to \pi^+\pi^-) = (-1.54\pm 0.29)\times 10^{-3},
\end{align}
in which the dominated direct $CP$ violation is defined by
\begin{align}
 A_{CP}^{\rm dir}(i\to f) \equiv \frac{|\mathcal{A}(i\to f)|^2-|\mathcal{A}(\overline i\to \overline f)|^2}{|\mathcal{A}(i\to f)|^2+|\mathcal{A}(\overline i\to \overline f)|^2} .
\end{align}
It is a milestone of particle physics, since $CP$ violation has been well established in the kaon and $B$ systems for many years \cite{Tanabashi:2018oca}, while the last piece of the puzzle, $CP$ violation in the charm sector, has not been observed until now. To find $CP$ violation in charm, many theoretical
and experimental devoted efforts were made in the past decade. On the other hand, with the discovery of doubly charmed baryon \cite{Aaij:2017ueg,Aaij:2018wzf,Aaij:2018gfl} and the progress of singly charmed baryon measurements \cite{Berger:2018pli,Zupanc:2013iki,Yang:2015ytm,Pal:2017ypp,Aaij:2017nsd,Aaij:2017xva,Ablikim:2018byv,Ablikim:2018jfs,Ablikim:2018woi,Ablikim:2018bir,Ablikim:2015prg,Ablikim:2015flg,Ablikim:2016tze,Ablikim:2016mcr,Ablikim:2016vqd,Ablikim:2017ors,Ablikim:2017iqd},
 plenty of theoretical interests focus on charmed baryon decays
\cite{Hsiao:2019yur,Geng:2019bfz,Shi:2019hbf,Aliev:2019lvd,Chen:2019ykv,Li:2019ekr,Shi:2019fph,Jiang:2018oak,Yu:2018com,Li:2017ndo,Yu:2017zst,Chen:2017sbg,Li:2017cfz,Wang:2017mqp,Meng:2017udf,Wang:2017azm,Karliner:2017qjm,Gutsche:2017hux,
Li:2017pxa,Guo:2017vcf,Xiao:2017udy,Lu:2017meb,Wang:2017qvg,Sharma:2017txj,Ma:2017nik,Meng:2017dni,Shi:2017dto,Hu:2017dzi,Wang:2018utj,
Zhao:2018mrg,Blin:2018pmj,Xing:2018lre,Bahtiyar:2018vub,Garcilazo:2018rwu,Wang:2018lhz,Park:2018wjk,Cheng:2018mwu,Dhir:2018twm,Geng:2018rse,He:2018joe,Zhao:2018mov,Grossman:2018ptn,Lu:2016ogy,Xie:2016evi,Faustov:2016yza,Li:2016qai,Xie:2017xwx,Geng:2017esc,Geng:2017mxn,Wang:2017gxe,Meinel:2017ggx,Geng:2018plk,
Cheng:2018hwl,Jiang:2018iqa,Zhao:2018zcb,Geng:2018bow,Faustov:2018dkn,Meinel:2016dqj,Geng:2018upx}.
However, only a few publications studied the $CP$ asymmetries in charmed baryon decays \cite{Kang:2010td,Grossman:2018ptn,Wang:2017gxe}.
The difference between $CP$ asymmetries of $\Lambda^+_c\to pK^+K^-$ and $\Lambda^+_c\to p\pi^+\pi^-$ modes has been measured by the LHCb Collaboration \cite{Aaij:2017xva} and no signal of $CP$ violation is found:
\begin{align}
\Delta A_{CP}^{\rm baryon} \equiv A_{CP}(\Lambda^+_c\to pK^+K^-)  - A_{CP}(\Lambda^+_c\to p\pi^+\pi^-) = (0.30\pm 0.91\pm 0.61)\%.
\end{align}

In charmed and bottomed meson decays, some relations for $CP$ asymmetries in (or beyond) the flavor $SU(3)$ limit are found \cite{Lipkin:1997ad,Fleischer:1999pa,Gronau:2000md,Gronau:2000zy,Nierste:2017cua,Yu:2017oky,Muller:2015rna,Grossman:2013lya,Grossman:2012ry,Grossman:2006jg,Pirtskhalava:2011va,Hiller:2012xm}.
For example, the direct $CP$ asymmetries in $D^0\to K^+K^-$ and $D^0\to \pi^+\pi^-$ decays have following relation  in the $U$-spin limit \cite{Grossman:2012ry,Grossman:2013lya}:
\begin{equation}\label{b8}
  A^{\rm dir}_{CP}(D^0\to K^+K^-) + A^{\rm dir}_{CP}(D^0\to \pi^+\pi^-) = 0.
\end{equation}
The two $CP$ asymmetries in $\Delta A_{CP}$ have opposite sign and hence are constructive in $\Delta A_{CP}$. But the two $CP$ asymmetries in $\Delta A_{CP}^{\rm baryon}$, as pointed out in \cite{Grossman:2018ptn}, do not have a relation like the ones in $\Delta A_{CP}$.
Prospects of measuring the $CP$ asymmetries of charmed baryon decays on LHCb \cite{Bediaga:2018lhg}, as well as Belle II \cite{Kou:2018nap}, are bright.
It is significative to study the relations for $CP$ asymmetries in the charmed baryon decays and then help to find some promising observables in experiments.

In Ref.~\cite{Grossman:2018ptn}, three $CP$ violation sum rules associated
with a complete interchange of $d$ and $s$ quarks are derived.
In this work, we illustrate that the complete interchange of $d$ and $s$ quarks
is a universal law to search for the $CP$ violation sum rules of two charmed hadron decay channels in the flavor $SU(3)$ limit.
With the universal law, hundreds of $CP$ violation sum rules can be found in the doubly and singly charmed baryon decays. The $CP$ violation sum rules could be tested in future measurements or provide a guide to find better observables for experiments. Besides, the branching fraction $\mathcal{B}r(\Xi^+_c\to pK^-\pi^+)$ and fragmentation-fraction ratio $f_{\Xi_b}/f_{\Lambda_b}$ are estimated in the $U$-spin limit.

The rest of this paper is organized as follows. In Sect.~\ref{hh}, the effective Hamiltonian of charm decay is decomposed into the $SU(3)$ irreducible representations. In Sect.~\ref{result1}, we derive the $CP$ violation sum rules for charmed meson and baryon decays and sum up a general law for $CP$ violation sum rules in charm.  In Sect.~\ref{result2}, we list some results of the $CP$ violation sum rules in charmed baryon decays.
Section~\ref{co} is a brief summary. The explicit $SU(3)$ decomposition of the operators in charm decays is presented in Appendix~\ref{dec}.

\section{Effective Hamiltonian of charm decay}\label{hh}
The effective
Hamiltonian in charm quark weak decay in the Standard Model (SM) can be written as \cite{Buchalla:1995vs}
 \begin{align}\label{hsm}
 \mathcal H_{\rm eff}={\frac{G_F}{\sqrt 2} }
 \left[\sum_{q=d,s}V_{cq_1}^*V_{uq_2}\left(\sum_{i=1}^2C_i(\mu)O_i(\mu)\right)
 -V_{cb}^*V_{ub}\left(\sum_{i=3}^6C_i(\mu)O_i(\mu)+C_{8g}(\mu)O_{8g}(\mu)\right)\right],
 \end{align}
 where $G_F$ is the Fermi coupling constant,
 $C_{i}$ is the Wilson coefficient of operator $O_i$.
The tree operators are
\begin{eqnarray}
O_1=(\bar{u}_{\alpha}q_{2\beta})_{V-A}
(\bar{q}_{1\beta}c_{\alpha})_{V-A},\quad
O_2=(\bar{u}_{\alpha}q_{2\alpha})_{V-A}
(\bar{q}_{1\beta}c_{\beta})_{V-A},
\end{eqnarray}
in which $\alpha,\beta$ are color indices, $q_{1,2}$ are $d$ and $s$
quarks.
The QCD penguin operators are
 \begin{align}
 O_3&=\sum_{q'=u,d,s}(\bar u_\alpha c_\alpha)_{V-A}(\bar q'_\beta
 q'_\beta)_{V-A},~~~
 O_4=\sum_{q'=u,d,s}(\bar u_\alpha c_\beta)_{V-A}(\bar q'_\beta q'_\alpha)_{V-A},
 \nonumber\\
 O_5&=\sum_{q'=u,d,s}(\bar u_\alpha c_\alpha)_{V-A}(\bar q'_\beta
 q'_\beta)_{V+A},~~~
 O_6=\sum_{q'=u,d,s}(\bar u_\alpha c_\beta)_{V-A}(\bar q'_\beta
 q'_\alpha)_{V+A},
 \end{align}
and the chromomagnetic-penguin operator is
\begin{eqnarray}
O_{8g}=\frac{g}{8\pi^2}m_c{\bar
u}\sigma_{\mu\nu}(1+\gamma_5)T^aG^{a\mu\nu}c.
\end{eqnarray}
The magnetic-penguin contributions can be included into the Wilson
coefficients for the penguin operators following the substitutions
\cite{Beneke:2003zv,Beneke:2000ry,Beneke:1999br}:
\begin{eqnarray}
C_{3,5}(\mu)\to& C_{3,5}(\mu) + \frac{\alpha_s(\mu)}{8\pi N_c}
\frac{2m_c^2}{\langle l^2\rangle}C_{8g}^{\rm eff}(\mu),\quad
C_{4,6}(\mu)\to& C_{4,6}(\mu) - \frac{\alpha_s(\mu)}{8\pi }
\frac{2m_c^2}{\langle l^2\rangle}C_{8g}^{\rm eff}(\mu),\label{mag}
\end{eqnarray}
with the effective Wilson coefficient $C_{8g}^{\rm eff}=C_{8g}+C_5$
and $\langle l^2\rangle$ being the averaged invariant mass squared of the virtual gluon emitted from the magnetic penguin operator.

The charm quark decays are categorized into three types, Cabibbo-favored(CF), singly Cabibbo-suppressed(SCS), and doubly Cabibbo-suppressed(DCS) decays, with the flavor structures of
\begin{align}
  c\to s\bar d u, \quad   c\to  d\bar d/ s \bar su,\quad c\to d\bar s u,
\end{align}
respectively. In the $SU(3)$ picture,
the operators in charm decays embed into the four-quark Hamiltonian,
\begin{align}\label{h}
  \mathcal{H}_{\rm eff}= \sum_{i,j,k=1}^{3}H^k_{ij}O^{ij}_k=\sum_{i,j,k=1}^{3}H^k_{ij}(\bar{q}^iq_k)(\bar{q}^jc).
\end{align}
Equation~\eqref{hsm} implies that the tensor
components of $H_{ij}^k$ can be obtained from the map $(\bar uq_1)(\bar q_2c)\rightarrow V^*_{cq_2}V_{uq_1}$ in current-current operators and $(\bar qq)(\bar uc)\rightarrow -V^*_{cb}V_{ub}$ in penguin operators and the others are zero.
 The non-zero components of the tensor $H_{ij}^k$ corresponding to tree operators in Eq.~\eqref{hsm} are
\begin{align}\label{ckm1}
 &H_{13}^2 = V_{cs}^*V_{ud},  \quad H^{2}_{12}=V_{cd}^*V_{ud},\quad H^{3}_{13}= V_{cs}^*V_{us}, \quad H^{3}_{12}=V_{cd}^*V_{us},
\end{align}
and the non-zero components of the tensor $H_{ij}^k$ corresponding to penguin operators in Eq.~\eqref{hsm} are
\begin{align}\label{ckm2}
 &H_{11}^1 = -V_{cb}^*V_{ub}, \quad H_{21}^2=-V_{cb}^*V_{ub}, \quad H_{31}^3=-V_{cb}^*V_{ub}.
\end{align}

The operator $O^{ij}_k$ is a representation of the $SU(3)$ group, which is decomposed as four irreducible representations: $\overline 3 \otimes  3 \otimes \overline3 =  \overline 3\oplus \overline 3^\prime\oplus  6 \oplus \overline{15}$. The explicit decomposition is \cite{Grossman:2012ry}
\begin{align}\label{hd}
  O_k^{ij}=\, \delta^j_k\Big(\frac{3}{8}O(\overline3)^i-\frac{1}{8}O(\overline3^{\prime})^i\Big)+
  \delta^i_k\Big(\frac{3}{8}O(\overline3^{\prime})^j-\frac{1}{8}O(\overline3)^j\Big)+\epsilon^{ijl}O( 6)_{lk}+O(\overline{15})_k^{ij}.
\end{align}
All components of the irreducible representations are listed in  Appendix \ref{dec}.
The non-zero components of $H_{ij}^k$ corresponding to tree operators in the $SU(3)$ decomposition are
\begin{align}\label{ckm3}
 &  H( 6)^{22}=-\frac{1}{2}V_{cs}^*V_{ud},\quad H( 6)^{23}=\frac{1}{4}(V_{cd}^*V_{ud}-V_{cs}^*V_{us}),  \quad H( 6)^{33}=  \frac{1}{2}V_{cd}^*V_{us},\nonumber \\
   &  H(\overline{15})^{1}_{11}=-\frac{1}{4}(V_{cd}^*V_{ud}+V_{cs}^*V_{us}) = \frac{1}{4}V_{cb}^*V_{ub}, \quad H(\overline{15})^{2}_{13}= \frac{1}{2}V_{cs}^*V_{ud},  \quad  H(\overline{15})^{3}_{12}=\frac{1}{2}V_{cd}^*V_{us},\nonumber \\
 &  H(\overline{15})^{2}_{12}= \frac{3}{8}V_{cd}^*V_{ud}-\frac{1}{8}V_{cs}^*V_{us},\quad H(\overline{15})^{3}_{13}=\frac{3}{8}V_{cs}^*V_{us}-\frac{1}{8}V_{cd}^*V_{ud}, \nonumber \\ & H(\overline 3)_1=V_{cd}^*V_{ud}+V_{cs}^*V_{us} = -V_{cb}^*V_{ub}.
\end{align}
The non-zero components of $H_{ij}^k$ corresponding to penguin operators in the $SU(3)$ decomposition are
\begin{align}\label{ckm4}
 H^P(\overline 3)_1=-V_{cb}^*V_{ub}, \quad H^P(\overline 3^\prime)_1=-3V_{cb}^*V_{ub}.
\end{align}
Here we use the superscript $P$ to differentiate penguin contributions from tree contributions. Equations~\eqref{ckm3} and \eqref{ckm4} were derived in \cite{Grossman:2012ry} for the first time. But the non-zero components $H(\overline{15})^{1}_{11}$ and $H(\overline 3)_1$ in the tree operator contributions are missing
in \cite{Grossman:2012ry}.

Recent studies for charmed baryon decays in the $SU(3)$ irreducible representation amplitude (IRA) approach \cite{Hsiao:2019yur,Geng:2018rse,He:2018joe,Geng:2018upx,Wang:2017azm,Shi:2017dto,Wang:2018utj,Lu:2016ogy,Geng:2017esc,Geng:2017mxn,Geng:2018plk,Geng:2018bow} do not analyze $CP$ asymmetries because they ignore the two 3-dimensional irreducible representations and make the approximation of $V^*_{cs}V_{us}\simeq -V^*_{cd}V_{ud}$ in the 15- and 6-dimensional irreducible representations, leading to the vanishing of the contributions proportional to $\lambda_b=V_{cb}^*V_{ub}$. If the contributions proportional to $\lambda_b$ are included, the $SU(3)$ irreducible representation amplitude approach then can be used to investigate $CP$ asymmetries in the charmed baryon decays.

\section{$CP$ violation sum rules in charmed meson/baryon decays}\label{result1}

In this section, we discuss the method to search for the relations for $CP$ asymmetries of charm decays in the flavor $SU(3)$ limit. We first analyze the $CP$ violation sum rules in charmed meson and baryon decays respectively, and then sum up a general law for the $CP$ violation sum rules in charm decays.

\subsection{$CP$ violation sum rules in charmed meson decays}

 Two $CP$ asymmetry sum rules for $D\to PP$ decays in the flavor $SU(3)$ limit have been given in \cite{Grossman:2012ry,Grossman:2013lya}:
\begin{align}\label{sur1}
&A^{\rm dir}_{CP}(D^0\to K^+K^-) + A^{\rm dir}_{CP}(D^0\to \pi^+\pi^-) = 0, \\\label{sur2}
&A^{\rm dir}_{C P}(D^+\to K^+\overline K^0) + A^{\rm dir}_{CP}(D^+_s\to \pi^+K^0)=0.
\end{align}
To see why the two sum rules are correct, we express the decay amplitudes of $D^0\to K^+K^-$, $D^0\to \pi^+\pi^-$, $D^+\to K^+\overline K^0$ and $D^+_s\to \pi^+K^0$ modes in the $SU(3)$ irreducible representation amplitude (IRA) approach.
The charmed meson anti-triplet is
\begin{align}
D^{i}= (D^0, D^+, D_s^+).
\end{align}
The pseudoscalar meson nonet is
\begin{eqnarray}
 P^i_j=  \left( \begin{array}{ccc}
   \frac{1}{\sqrt 2} \pi^0+  \frac{1}{\sqrt 6} \eta_8    & \pi^+  & K^+ \\
    \pi^- &   - \frac{1}{\sqrt 2} \pi^0+ \frac{1}{\sqrt 6} \eta_8   & K^0 \\
    K^- & \overline K^0 & -\sqrt{2/3}\eta_8 \\
  \end{array}\right) +  \frac{1}{\sqrt 3} \left( \begin{array}{ccc}
   \eta_1    & 0  & 0 \\
    0 &  \eta_1   & 0 \\
   0 & 0 & \eta_1 \\
  \end{array}\right).
\end{eqnarray}
To obtain the $SU(3)$ irreducible representation amplitude of $D\to PP$ decay, one takes various representations in Eqs.~\eqref{ckm3} and \eqref{ckm4} and contracts all indices in $D^i$ and light meson $P^i_j$ with various combinations:
\begin{align}\label{amp1}
{\cal A}^{\rm tree}_{D\to PP} = &\,a_{15}D^i H(\overline{15})_{ij}^kP^j_lP^l_k + b_{15}D^i H(\overline{15})_{ij}^kP^j_kP^l_l + c_{15}D^i H(\overline{15})_{jl}^kP^j_iP^l_k
 \nonumber\\
  & + a_6D^i H( 6)_{ij}^kP^j_lP^l_k + b_6D^i H(6)_{ij}^kP^j_kP^l_l + c_6D^i H(6)_{jl}^kP^j_iP^l_k\nonumber \\&+a_3 D^i H(\overline3)_i P_k^jP_j^k +b_3 D^iH(\overline3)_i P_k^kP_j^j+c_3 D^i H(\overline3)_k P_i^kP_j^j\nonumber\\
  &+d_3 D^i H(\overline3)_k P_i^jP_j^k.
\end{align}
\begin{align}\label{amp2}
{\cal A}^{\rm penguin}_{D\to PP} =
&\,Pa_3 D^i H^P(\overline3)_i P_k^jP_j^k +Pb_3 D^i H^P(\overline3)_i P_k^kP_j^j+Pc_3 D^i H^P(\overline3)_k P_i^kP_j^j\nonumber\\
  &+Pd_3 D^i H^P(\overline3)_k P_i^jP_j^k+Pa^\prime_{3} D^i H^P(\overline3^\prime)_i P_k^jP_j^k +Pb^\prime_{3} D^i H^P(\overline3^\prime)_i P_k^kP_j^j\nonumber\\&+Pc^\prime_{3} D^i H^P(\overline3^\prime)_k P_i^kP_j^j+Pd^\prime_{3} D^i H^P(\overline3^\prime)_k P_i^jP_j^k.
\end{align}
Notice that only the first components of $\overline 3$ and $\overline 3^{\prime}$ irreducible representations are non-zero. Some amplitudes, for example, $a_3$, $Pa_3$ and $Pa^\prime_{3}$ are always appear simultaneously since they correspond to the same contraction. Noting that $H^P(\overline 3^\prime)_1 = 3 H(\overline 3)_1= 3 H^P(\overline 3)_1 =-3V_{cb}^*V_{ub}$, if we define
\begin{align}
Pa=a_3 + Pa_3+ 3Pa^\prime_{3},\quad Pb=b_3 + Pb_3+ 3Pb^\prime_{3},\quad  \ldots,
\end{align}
the amplitude of $D\to PP$ decay will be reduced to
\begin{align}\label{amp3}
{\cal A}^{\rm tree+penguin}_{D\to PP} = &\,a_{15}D^i H(\overline{15})_{ij}^kP^j_lP^l_k + b_{15}D^i H(\overline{15})_{ij}^kP^j_kP^l_l + c_{15}D^i H(\overline{15})_{jl}^kP^j_iP^l_k
 \nonumber\\
  & + a_6D^i H( 6)_{ij}^kP^j_lP^l_k + b_6D^i H(6)_{ij}^kP^j_kP^l_l + c_6D^i H(6)_{jl}^kP^j_iP^l_k\nonumber \\&+Pa D^i H(\overline3)_i P_k^jP_j^k +Pb D^iH(\overline3)_i P_k^kP_j^j+Pc D^i H(\overline3)_k P_i^kP_j^j\nonumber\\
  &+Pd D^i H(\overline3)_k P_i^jP_j^k.
\end{align}
With Eq.~\eqref{amp3},
the decay amplitudes of the $D^0\to K^+K^-$, $D^0\to \pi^+\pi^-$, $D^+\to K^+\overline K^0$ and $D^+_s\to \pi^+K^0$ modes read
\begin{align}\label{b1}
  \mathcal{A}(D^0\to K^+K^-) & = -\lambda_d(\frac{1}{8}a_{15}+\frac{1}{8}c_{15}+\frac{1}{4}a_6-\frac{1}{4}c_6)
  +\lambda_s(\frac{3}{8}a_{15}+\frac{3}{8}c_{15}+\frac{1}{4}a_6-\frac{1}{4}c_6)\nonumber\\
  &~~~~~~~~~~~~~~-\lambda_b(2Pa+Pd- a_{15}/4),\\\label{b2}
   \mathcal{A}(D^0\to \pi^+\pi^-) & =   \lambda_d(\frac{3}{8}a_{15}+\frac{3}{8}c_{15}+\frac{1}{4}a_6-\frac{1}{4}c_6)
  -\lambda_s(\frac{1}{8}a_{15}+\frac{1}{8}c_{15}+\frac{1}{4}a_6-\frac{1}{4}c_6)\nonumber\\
  &~~~~~~~~~~~~~~-\lambda_b(2Pa+Pd- a_{15}/4),\\\label{b3}
   \mathcal{A}(D^+\to K^+\overline K^0) & =   \lambda_d(\frac{3}{8}a_{15}-\frac{1}{8}c_{15}-\frac{1}{4}a_6+\frac{1}{4}c_6)
  -\lambda_s(\frac{1}{8}a_{15}-\frac{3}{8}c_{15}-\frac{1}{4}a_6+\frac{1}{4}c_6)\nonumber\\
  &~~~~~~~~~~~~~~-\lambda_bPd,\\\label{b4}
   \mathcal{A}(D^+_s\to \pi^+K^0) & =   -\lambda_d(\frac{1}{8}a_{15}-\frac{3}{8}c_{15}-\frac{1}{4}a_6+\frac{1}{4}c_6)
  +\lambda_s(\frac{3}{8}a_{15}-\frac{1}{8}c_{15}-\frac{1}{4}a_6+\frac{1}{4}c_6)\nonumber\\
  &~~~~~~~~~~~~~~-\lambda_bPd,
\end{align}
'in which $\lambda_d = V^*_{cd}V_{ud}$, $\lambda_s = V^*_{cs}V_{us}$, $\lambda_b = V^*_{cb}V_{ub}$.
Equations~\eqref{b1}-\eqref{b4} are consistent with \cite{Grossman:2012ry} except for the last terms in Eqs.~\eqref{b1} and \eqref{b2} because of the non-vanishing $H(\overline{15})^{1}_{11}$ component in Eq.~\eqref{ckm3}.
From above formulas, the $CP$ violation sum rules listed in Eqs.~\eqref{sur1} and \eqref{sur2} are derived if  the approximation of
\begin{align}\label{app}
\lambda_b\lambda_d= -\lambda_b(\lambda_s+\lambda_b)=-(\lambda_b\lambda_s+\lambda^2_b)\simeq -\lambda_b\lambda_s
\end{align}
is used.
Besides,
the decay amplitude of $D^0\to K^0\overline K^0$ is expressed as
\begin{align}\label{b5}
\mathcal{A}(D^0\to K^0\overline K^0) = -\lambda_b(2Pa+\frac{1}{4}a_{15}).
\end{align}
The direct $CP$ asymmetry in $D^0\to K^0\overline K^0$ decay is zero in the flavor $SU(3)$ limit:
\begin{align}\label{sur3}
&A^{\rm dir}_{CP}(D^0\to K^0\overline K^0) = 0.
\end{align}

For the $CP$ violation relations \eqref{sur1} and \eqref{sur2}, the decay amplitudes of two channels are connected by the interchange of $\lambda_d \leftrightarrow \lambda_s$, and their initial and final states are connected by the interchange of $d \leftrightarrow s$:
\begin{align}
D^+ \leftrightarrow D^+_s,\quad D^0 \leftrightarrow D^0,\quad K^+ \leftrightarrow \pi^+, \quad K^- \leftrightarrow \pi^-,  \quad K^0\leftrightarrow \overline K^0.
\end{align}
For $D^0\to K^0\overline K^0$ decay, its corresponding mode in the interchange of $d \leftrightarrow s$ is itself.
So all $CP$ violation relations in Eqs.~\eqref{sur1}, \eqref{sur2} and \eqref{sur3} are associated with $U$-spin transformation.

On the other hand, Eqs.~\eqref{sur1}, \eqref{sur2} and \eqref{sur3} include all the SCS modes without $\pi^0$, $\eta^{(\prime)}$ in the final states in $D\to PP$ decays.
Mesons $\pi^0$ and $\eta^{(\prime)}$ do not have definite $U$-spin quantum numbers.
Under the interchange of $d\leftrightarrow s$, there are no mesons corresponding to $\pi^0$ and $\eta^{(\prime)}$. For example, $\pi^0$ has the quark constituent of $(\bar d d- \bar uu)/\sqrt{2}$.  Under the interchange of $d\leftrightarrow s$, $(\bar d d- \bar uu)/\sqrt{2}$ turns into $(\bar s s- \bar uu)/\sqrt{2}$. No meson has the  quark constituent of $(\bar s s- \bar uu)/\sqrt{2}$.
So those decay channels involving $\pi^0$, $\eta^{(\prime)}$ do not have their corresponding modes in the interchange of $d\leftrightarrow s$, and then have no simple $CP$ violation sum rules with two channels.

In fact, not only the $D\to PP$ decays, there are also some $CP$ violation sum rules in the $D\to PV$ decays \cite{Grossman:2013lya}
\begin{align}\label{b6}
&A_{CP}^{\rm dir}(D^0\to \pi^-\rho^+) + A_{CP}^{\rm dir}(D^0\to K^-K^{*+}) = 0, \\
&A_{CP}^{\rm dir}(D^0\to \pi^+\rho^-) + A_{CP}^{\rm dir}(D^0\to K^+K^{*-})=0,\\
&A_{CP}^{\rm dir}(D^+\to \overline K^0K^{*+}) + A_{CP}^{\rm dir}(D^+_s\to K^0\rho^+)=0,\\
&A_{CP}^{\rm dir}(D^+\to K^+\overline K^{*0}) + A_{CP}^{\rm dir}(D^+_s\to \pi^+K^{*0})=0,\\ \label{b7}
&A_{CP}^{\rm dir}(D^0\to K^0\overline K^{*0}) + A_{CP}^{\rm dir}(D^0\to \overline K^0K^{*0})=0.
\end{align}
The detailed derivation of these sum rules is similar to $D\to PP$ and can be found in Ref.~\cite{Grossman:2012ry}.
Again, all the $CP$ violation sum rules in $D\to PV$ decays are associated with a complete interchange of $d$ and $s$
quarks, and all the singly Cabibbo-suppressed $D\to PV$ modes with all final states having definite $U$-spin quantum numbers are included in Eqs.~\eqref{b6}-\eqref{b7}.

\subsection{$CP$ violation sum rules in charmed baryon decays}
\begin{table*}
\caption{$SU(3)$ irreducible representation amplitudes in $\mathcal{B}_{c\overline 3}\to \mathcal{B}_{10}M$ decays, in which only those modes that all initial and final states have definite $U$-spin quantum numbers are listed.  }\label{ampsu3}
%\resizebox{\textwidth}{!}{ %
%\begin{ruledtabular}
\begin{tabular}{|c|c|}
\hline
 Channel & Amplitude \\\hline
 ~~~$\Lambda^+_c\to \Delta^0\pi^+$~~~&  ~~~~~~$\frac{1}{8\sqrt{3}}\lambda_d(6 e_1-6 e_2+5 e_3-2 e_4)-\frac{1}{8\sqrt{3}}\lambda_s(2 e_1-2 e_2-e_3-2 e_4)+\frac{1}{\sqrt{3}}\lambda_b Pe$ ~~~~~~ \\\hline
 $\Lambda^+_c\to \Sigma^{*+}K^0$& $\frac{1}{8\sqrt{3}}\lambda_d(2 e_1-2 e_2+3 e_3+2 e_4)+\frac{1}{8\sqrt{3}}\lambda_s(2 e_1+6 e_2-e_3-2 e_4)-\frac{1}{\sqrt{3}}\lambda_bPe$  \\\hline
 $\Lambda^+_c\to \Sigma^{*0}K^+$&  $\frac{1}{8\sqrt{6}}\lambda_d(6 e_1+2 e_2+5 e_3-2 e_4)-\frac{1}{8\sqrt{6}}\lambda_s(2 e_1+6 e_2-e_3-2 e_4)+\frac{1}{\sqrt{6}}\lambda_bPe$ \\\hline
 $\Lambda^+_c\to \Delta^{++}\pi^-$&  $\frac{1}{8}\lambda_d(2 e_1-2 e_2+3 e_3+2 e_4)+\frac{1}{8}\lambda_s(2 e_1-2 e_2-e_3-2 e_4)-\lambda_bPe$ \\\hline
 $\Xi^+_c\to \Xi^{*0}K^+$&  $-\frac{1}{8\sqrt{3}}\lambda_d(2 e_1-2 e_2-e_3-2 e_4)+\frac{1}{8\sqrt{3}}\lambda_s(6 e_1-6 e_2+5 e_3-2 e_4)+\frac{1}{\sqrt{3}}\lambda_bPe$ \\\hline
 $\Xi^+_c\to \Delta^{+}\overline K^0$&  $\frac{1}{8\sqrt{3}}\lambda_d(2 e_1+6 e_2-e_3-2 e_4)+\frac{1}{8\sqrt{3}}\lambda_s(2 e_1-2 e_2+3 e_3+2 e_4)-\frac{1}{\sqrt{3}}\lambda_bPe$ \\\hline
 $\Xi^+_c\to \Sigma^{*0}\pi^+$& $-\frac{1}{8\sqrt{6}}\lambda_d(2 e_1+6 e_2-e_3-2 e_4)+\frac{1}{8\sqrt{6}}\lambda_s (6 e_1+2 e_2+5 e_3-2 e_4)+\frac{1}{\sqrt{6}}\lambda_bPe$ \\\hline
 $\Xi^+_c\to \Delta^{++}K^-$&   $\frac{1}{8}\lambda_d(2 e_1-2 e_2-e_3-2 e_4)+\frac{1}{8}\lambda_s(2 e_1-2 e_2+3 e_3+2 e_4)-\lambda_bPe$\\\hline
 $\Xi^0_c\to \Sigma^{*-}\pi^+$& $-\frac{1}{2\sqrt{3}}\lambda_d(e_3-e_4)+\frac{1}{2\sqrt{3}}\lambda_s(e_3-e_4)$ \\\hline
 $\Xi^0_c\to \Xi^{*-}K^+$& $-\frac{1}{2\sqrt{3}}\lambda_d(e_3-e_4)+\frac{1}{2\sqrt{3}}\lambda_s(e_3-e_4)$ \\\hline
 $\Xi^0_c\to \Delta^{0}\overline K^0$& $-\frac{1}{8\sqrt{3}}\lambda_d(6 e_1-6 e_2+e_3+2 e_4)+\frac{1}{8\sqrt{3}}\lambda_s(2 e_1-2 e_2+3 e_3+2 e_4)-\frac{1}{\sqrt{3}}\lambda_bPe$ \\\hline
 $\Xi^0_c\to \Xi^{*0}K^0$& $-\frac{1}{8\sqrt{3}}\lambda_d(2 e_1-2 e_2+3 e_3+2 e_4)+\frac{1}{8\sqrt{3}}\lambda_s(6 e_1-6 e_2+e_3+2 e_4)+\frac{1}{\sqrt{3}}\lambda_bPe$ \\\hline
 $\Xi^0_c\to \Sigma^{*+}\pi^-$&   $-\frac{1}{8\sqrt{3}}\lambda_d(2 e_1-2 e_2+3 e_3+2 e_4)+\frac{1}{8\sqrt{3}}\lambda_s(6 e_1+2 e_2+e_3+2 e_4)+\frac{1}{\sqrt{3}}\lambda_bPe$ \\\hline

 $\Xi^0_c\to \Delta^{+}K^-$&  $-\frac{1}{8\sqrt{3}}\lambda_d(6 e_1+2 e_2+e_3+2 e_4)+\frac{1}{8\sqrt{3}}\lambda_s(2 e_1-2 e_2+3 e_3+2 e_4)-\frac{1}{\sqrt{3}}\lambda_bPe$  \\\hline
\end{tabular}
%\end{ruledtabular}
\end{table*}

In this subsection, we take charmed baryon decays into one pseudoscalar meson and one decuplet baryon as examples to show the complete interchange of $d \leftrightarrow s$ is still valid for the $CP$ violation sum rules in charmed baryon decays.
The charmed anti-triplet baryon is expressed as
\begin{eqnarray}
 \mathcal{B}_{c\overline 3}=  \left( \begin{array}{ccc}
   0   & \Lambda_c^+  & \Xi_c^+ \\
    -\Lambda_c^+ &   0   & \Xi_c^0 \\
    -\Xi_c^+ & -\Xi_c^0 & 0 \\
  \end{array}\right).
\end{eqnarray}
The light baryon decuplet is given as
\begin{align}
 &\Delta^{++}  =  \mathcal{B}_{10}^{111},  \quad \Delta^{-}= \mathcal{B}_{10}^{222},\quad  \Omega^-= \mathcal{B}_{10}^{333}, \nonumber\\
 & \Delta^{+} = \frac{1}{\sqrt{3}}( \mathcal{B}_{10}^{112} +  \mathcal{B}_{10}^{121} + \mathcal{B}_{10}^{211}),\quad  \Delta^{0} = \frac{1}{\sqrt{3}}( \mathcal{B}_{10}^{122} +  \mathcal{B}_{10}^{212} +  \mathcal{B}_{10}^{221}),\nonumber\\
&\Sigma^{*+}= \frac{1}{\sqrt{3}}( \mathcal{B}_{10}^{113} + \mathcal{B}_{10}^{131} + \mathcal{B}_{10}^{311}), \quad \Sigma^{*-}=\frac{1}{\sqrt{3}} ( \mathcal{B}_{10}^{223} + \mathcal{B}_{10}^{232} + \mathcal{B}_{10}^{322}), \nonumber\\ &
 \Xi^{*0}=\frac{1}{\sqrt{3}}( \mathcal{B}_{10}^{133} + \mathcal{B}_{10}^{313} + \mathcal{B}_{10}^{331}),\quad
 \Xi^{*-}=\frac{1}{\sqrt{3}}( \mathcal{B}_{10}^{233} + \mathcal{B}_{10}^{323}+ \mathcal{B}_{10}^{332}),\nonumber\\
 & \Sigma^{*0}=\frac{1}{\sqrt{6}}( \mathcal{B}_{10}^{123} +  \mathcal{B}_{10}^{132} + \mathcal{B}_{10}^{213}+  \mathcal{B}_{10}^{231} + \mathcal{B}_{10}^{312} + \mathcal{B}_{10}^{321}).
\end{align}
The $SU(3)$ irreducible representation amplitude of $\mathcal{B}_{c\overline 3}\to \mathcal{B}_{10} M$ decay can be written as
\begin{align}
  \mathcal{A}^{\rm tree}_{\mathcal{B}_{c\overline 3}\to \mathcal{B}_{10}M} =
  & \,e_1(\mathcal{B}_{c\overline3})_{ij}H(\overline{15})^j_{kl}M^i_m\overline {\mathcal{B}}_{10}^{klm}+e_2(\mathcal{B}_{c\overline3})_{ij}H(\overline{15})^k_{lm}M^j_k\overline {\mathcal{B}}_{10}^{ilm}
  +e_3({\mathcal{B}}_{c\overline3})_{ij}H(\overline{15})^j_{kl}M^l_m\overline {\mathcal{B}}_{10}^{ikm}\nonumber\\
&+  e_4({\mathcal{B}}_{c\overline3})_{ij}H( 6)^j_{kl}M^l_m\overline {\mathcal{B}}_{10}^{ikm} +e_5({\mathcal{B}}_{c\overline3})_{ij}H(\overline{15})^j_{kl}M^m_m\overline {\mathcal{B}}_{10}^{ikl}\nonumber\\
&+e_6({\mathcal{B}}_{c\overline3})_{ij}H(\overline 3)_{k}M^j_m\overline {\mathcal{B}}_{10}^{ikm},
\end{align}
\begin{align}
  \mathcal{A}^{\rm penguin}_{\mathcal{B}_{c\overline 3}\to \mathcal{B}_{10}M} =
  \,Pe_6({\mathcal{B}}_{c\overline3})_{ij}H^P(\overline 3)_{k}M^j_m\overline {\mathcal{B}}_{10}^{ikm}
+Pe_7({\mathcal{B}}_{c\overline3})_{ij}H^P(\overline 3^\prime)_{k}M^j_m\overline {\mathcal{B}}_{10}^{ikm}.
\end{align}
Similar to $D\to PP$ decay, if we define
\begin{align}
Pe=e_6 + Pe_6+ 3Pe_7,
\end{align}
the amplitude of $\mathcal{B}_{c\overline 3}\to \mathcal{B}_{10} M$ decay will be reduced to
\begin{align}\label{ira}
  \mathcal{A}^{\rm tree+penguin}_{\mathcal{B}_{c\overline 3}\to \mathcal{B}_{10}M} =
  & \,e_1(\mathcal{B}_{c\overline3})_{ij}H(\overline{15})^j_{kl}M^i_m\overline {\mathcal{B}}_{10}^{klm}+e_2(\mathcal{B}_{c\overline3})_{ij}H(\overline{15})^k_{lm}M^j_k\overline {\mathcal{B}}_{10}^{ilm}
  +e_3({\mathcal{B}}_{c\overline3})_{ij}H(\overline{15})^j_{kl}M^l_m\overline {\mathcal{B}}_{10}^{ikm}\nonumber\\
&+  e_4({\mathcal{B}}_{c\overline3})_{ij}H( 6)^j_{kl}M^l_m\overline {\mathcal{B}}_{10}^{ikm} +e_5({\mathcal{B}}_{c\overline3})_{ij}H(\overline{15})^j_{kl}M^m_m\overline {\mathcal{B}}_{10}^{ikl}\nonumber\\
&+Pe({\mathcal{B}}_{c\overline3})_{ij}H(\overline 3)_{k}M^j_m\overline {\mathcal{B}}_{10}^{ikm}.
\end{align}
The first four terms are the same with the formula given in \cite{Savage:1989qr}. The fifth term is the decay amplitude associated with singlet $\eta_1$, and the six term is the amplitude proportional to $\lambda_b$.
With Eq.~\eqref{ira}, the $SU(3)$ irreducible representation amplitudes of $\mathcal{B}_{c\overline 3}\to \mathcal{B}_{10}M$ decays are obtained. The results are listed in Table~\ref{ampsu3}.

From Table~\ref{ampsu3}, seven $CP$ violation sum rules in the $SU(3)_F$ limit for the charmed baryon decays into one pseudoscalar meson and one decuplet baryon are found:
\begin{align}\label{b11}
 & A^{\rm dir}_{CP}(\Lambda_{c}^{+}\to\Delta^0\pi^+) +  A^{\rm dir}_{CP}(\Xi_{c}^{+}\to \Xi^{*0}K^+)=0,\\
  &  A^{\rm dir}_{CP}(\Lambda_{c}^{+}\to \Sigma^{*+}K^0) +  A^{\rm dir}_{CP}(\Xi_{c}^{+}\to \Delta^{+}\overline K^0)=0,\\
  &  A^{\rm dir}_{CP}(\Lambda_{c}^{+}\to \Sigma^{*0}K^+) +  A^{\rm dir}_{CP}(\Xi_{c}^{+}\to \Sigma^{*0}K^+)=0,\\
  &  A^{\rm dir}_{CP}(\Lambda_{c}^{+}\to \Delta^{++}\pi^-) +  A^{\rm dir}_{CP}(\Xi_{c}^{+}\to \Delta^{++} K^-)=0,\\
  &  A^{\rm dir}_{CP}(\Xi_{c}^{0}\to \Sigma^{*-}\pi^+) +  A^{\rm dir}_{CP}(\Xi_{c}^{0}\to \Xi^{*-}K^+)=0,\\
  &  A^{\rm dir}_{CP}(\Xi_{c}^{0}\to \Delta^{0}\overline K^0) +  A^{\rm dir}_{CP}(\Xi_{c}^{0}\to \Xi^{*0} K^0)=0,\\\label{b12}
  &  A^{\rm dir}_{CP}(\Xi_{c}^{0}\to \Sigma^{*+}\pi^-) +  A^{\rm dir}_{CP}(\Xi_{c}^{0}\to \Delta^{+}K^-)=0.
\end{align}
Similar to the charmed meson decays, all the $CP$ violation sum rules are associated with a complete interchange of $d$ and $s$ quarks in the initial and final states.
For charmed anti-triplet baryons,
\begin{align}
\Lambda^+_c\leftrightarrow \Xi^+_c,\quad \Xi^0_c\leftrightarrow \Xi^0_c.
\end{align}
For light decuplet baryons,
\begin{align}
\Delta^0\leftrightarrow \Xi^{*0},\quad \Sigma^{*+}\leftrightarrow \Delta^+,\quad \Sigma^{*0}\leftrightarrow \Sigma^{*0}, \quad \Delta^{++}\leftrightarrow \Delta^{++},\quad \Xi^{*-}\leftrightarrow \Sigma^{*-},\quad \Delta^{-}\leftrightarrow \Omega^{-}.
\end{align}
Also, Eqs.~\eqref{b11}-\eqref{b12} include all the SCS modes with all associated particles having definite $U$-spin quantum numbers in $\mathcal{B}_{c\overline 3}\to \mathcal{B}_{10}M$ decays.

For other types of charm baryon decay, for example $\mathcal{B}_{c\overline 3}\to \mathcal{B}_{8}M$ decay, multi-body decay and doubly charmed baryon decay, the treatments of their $SU(3)$ irreducible representation amplitudes are similar to $\mathcal{B}_{c\overline 3}\to \mathcal{B}_{10}M$.
Related discussions can be found in Refs.~\cite{Wang:2018utj,Shi:2017dto,Wang:2017azm,Savage:1989qr}. But notice that the contributions proportional to $\lambda_b$ are neglected in this literature. To get a complete expression of decay amplitude and then analyze the $CP$ asymmetries, the neglected terms must be found back, just like we have done in this work. One can check that the $CP$ violation sum rules associated with the complete interchange of $d$ and $s$ quarks works in various types of decay.

\subsection{A universal law for $CP$ violation sum rules in the charm sector}

From the above discussions, one can find the $CP$ violation sum rules in the $SU(3)_F$ limit are always associated with a complete interchange of $d$ and $s$ quarks. In this subsection, we illustrate that it is a universal law in the charm sector.

Firstly, the complete interchange of $d \leftrightarrow s$ quarks in initial and final states leads to the interchange of $d\leftrightarrow s$ in operators $O^{ij}_k$.
It can be understood in following argument.
In the IRA approach, each decay amplitude connects to one invariant tensor [in which all covariant indices are contracted with contravariant indices; see Eq.~\eqref{amp3} for example],
no matter charm meson or baryon decays and two- or multi-body decays.
If a complete interchange of $d$ and $s$ quarks is performed in the tensors corresponding to initial and final states, the complete interchange of $d \leftrightarrow s$ must be performed in tensor $H_{ij}^{k}$
in order to keep all covariant and contravariant indices contracted. From Eq.~\eqref{h}, one can find $H_{ij}^{k}$ corresponds to $O^{ij}_{k}$ one by one.
So the $d$ and $s$ quark constituents in operators $O^{ij}_k$ must be interchanged.
In physics, if the quark constituents of all initial and final particles in one decay channel are replaced by $d \to s$ and $s \to d$, the quark constituents in the effective weak vertexes should be replaced by $d \to s$ and $s \to d$ also. The operators $O^{ij}_k$ are abstracted from the effective weak vertexes, so the quark constituents of operators $O^{ij}_k$ transform as a complete interchange of $d \leftrightarrow s$.

Secondly,
the interchange of $d\leftrightarrow s$ in operators $O^{ij}_k$ leads to the
 decay amplitudes proportional to $\lambda_d/\lambda_s$
are connected by the interchange of $\lambda_d \leftrightarrow \lambda_s$ and the decay amplitudes proportional to $\lambda_b$ are the same in the flavor $SU(3)$ symmetry.
The contributions proportional to $\lambda_d/\lambda_s$ in SCS decays are induced by following operators in the $SU(3)$ irreducible representation:
\begin{align}
O(6)_{23}, \quad O(\overline {15})^{12}_2,\quad  O(\overline {15})^{13}_3.
\end{align}
Under the interchange of $d \leftrightarrow s$, these operators are transformed as
\begin{align}
O(6)_{23}\,\,\leftrightarrow\,\,  -O(6)_{23} , \quad O(\overline {15})^{12}_2\,\, \leftrightarrow  \,\, O(\overline {15})^{13}_3.
\end{align}
These properties can be read from the explicit $SU(3)$ decomposition of $O^{ij}_k$; see Appendix \ref{dec}.
%This conclusion is understandable since operators $O(6)_{23}$, $O(\overline {15})^{12}_2$ and $O(\overline {15})^{13}_3$ arise from operators $O^{12}_2$ and $O^{13}_3$, and they transform as
%$ O^{12}_2\,\leftrightarrow  \, O^{13}_3$
%under  the  interchange of $d \leftrightarrow s$.
The corresponding CKM matrix elements then transform as
\begin{align}\label{b13}
H(6)^{23}\,\,\leftrightarrow\,\,  -H(6)^{23} , \quad H(\overline {15})_{12}^2\,\, \leftrightarrow  \,\, H(\overline {15})_{13}^3.
\end{align}
According to Eq.~\eqref{ckm3}, Eq.~\eqref{b13} equals
\begin{align}\label{ex}
\frac{1}{4}(\lambda_d-\lambda_s)\,\,\leftrightarrow\,\, - \frac{1}{4}(\lambda_d-\lambda_s),\quad \frac{3}{8}\lambda_d-\frac{1}{8}\lambda_s\,\, \leftrightarrow  \,\, \frac{3}{8}\lambda_s-\frac{1}{8}\lambda_d.
\end{align}
One can find that Eq.~\eqref{ex} is equivalent to the interchange of $\lambda_d\leftrightarrow \lambda_s$.
The contributions  proportional to $\lambda_b$ in the SCS decays are induced by the following operators:
\begin{align}
O(\overline {15})^{11}_1, \quad O(\overline {3})^{1},\quad  O(\overline {3}^\prime)^{1}.
\end{align}
Form Appendix \ref{dec}, it is found that
these operators are invariable
under the  interchange of $d \leftrightarrow s$, as are the corresponding CKM matrix elements.

Thirdly, if two decay channels have the relations that their decay amplitudes proportional to $\lambda_d/\lambda_s$
are connected by the interchange of $\lambda_d \leftrightarrow \lambda_s$ and the decay amplitudes proportional to $\lambda_b$ are the same, the sum of their direct $CP$ asymmetries is zero in the $SU(3)_F$ limit under the approximation in Eq.~\eqref{app}. For one decay mode with amplitude of
\begin{align}\label{x1}
  \mathcal{A}(i\to f)= &\,\lambda_d A + \lambda_s B +\lambda_b C \nonumber\\
   & =-(\lambda_s+|\lambda_b|\, e^{i\phi}) \,|A|\,e^{i\delta_A} + \lambda_s \,|B|\,e^{i\delta_B} +|\lambda_b|\,e^{i\phi} \,|C|\,e^{i\delta_C},
\end{align}
its $CP$ asymmetry in the order of $\mathcal{O}(\lambda_b)$ is derived as
\begin{align}
A^{\rm dir}_{CP}(i\to f) & = \frac{|\lambda_d A + \lambda_s B +\lambda_b C|^2-|\lambda^*_d A + \lambda^*_s B +\lambda^*_b C|^2}{|\lambda_d A + \lambda_s B +\lambda_b C|^2+|\lambda^*_d A + \lambda^*_s B +\lambda^*_b C|^2}\nonumber\\
&\simeq 2 \frac{|\lambda_b|}{\lambda_s}\frac{|AB|\sin(\delta_A-\delta_B)-|AC|\sin(\delta_A-\delta_C)+|BC|\sin(\delta_B-\delta_C)}
{|A|^2+|B|^2-2|AB|\cos(\delta_A-\delta_B)}\sin\phi.
\end{align}
For one decay mode with amplitude of
\begin{align}\label{x2}
  \mathcal{A}(i^\prime\to f^\prime)&= \lambda_d B + \lambda_s A +\lambda_b C,
\end{align}
which is connected to Eq.~\eqref{x1} by $\lambda_d\leftrightarrow \lambda_s$,
its $CP$ asymmetry in the order of $\mathcal{O}(\lambda_b)$ is derived as
\begin{align}
A^{\rm dir}_{CP}(i^\prime\to f^\prime) \simeq -2 \frac{|\lambda_b|}{\lambda_s}\frac{|AB|\sin(\delta_A-\delta_B)-|AC|\sin(\delta_A-\delta_C)+|BC|\sin(\delta_B-\delta_C)}
{|A|^2+|B|^2-2|AB|\cos(\delta_A-\delta_B)}\sin\phi.
\end{align}
It is apparent that
\begin{align}
A^{\rm dir}_{CP}(i\to f)+A^{\rm dir}_{CP}(i^\prime\to f^\prime)\simeq 0.
\end{align}

Based on the above analysis, a useful method to search for the $CP$ violation sum rules with two charmed hadron decay channels is proposed:
\begin{itemize}
  \item For one type of charmed hadron decay, write down all the SCS decay modes in which the associated hadrons have definite $U$-spin quantum numbers;
  \item For each decay mode, find the corresponding decay mode in the complete interchange of $d\leftrightarrow s$;
  \item If there are two decay modes connected by the interchange of $d\leftrightarrow s$, the sum of their direct $CP$ asymmetries is zero in the $SU(3)_F$ limit;
 \item  If the corresponding decay mode is itself, the direct $CP$ asymmetry in this mode is zero in the $SU(3)_F$ limit.
\end{itemize}

\section{Results and discussions}\label{result2}
With the method proposed in Sect. \ref{result1}, one can find many sum rules for $CP$ asymmetries in charm meson/baryon decays.
There are hundreds of sum rules for $CP$ asymmetries in the singly and doubly charmed baryon decays.
We are not going to list all the $CP$ violation sum rules, but only present some of them as examples.

Under the complete interchange of $d \leftrightarrow s$,
the light octet baryons are interchanged as
\begin{align}
p\leftrightarrow \Sigma^+,\quad n\leftrightarrow \Xi^0,\quad \Sigma^-\leftrightarrow \Xi^-.
\end{align}
The sum rules for $CP$ asymmetries in charmed baryon decays into one pseudoscalar meson and one octet baryon are
\begin{align}
 & A^{\rm dir}_{CP}(\Lambda_{c}^{+}\to\Sigma^+K^0) +  A^{\rm dir}_{CP}(\Xi_{c}^{+}\to p\overline K^0)=0,\\
  &  A^{\rm dir}_{CP}(\Lambda_{c}^{+}\to n\pi^+) +  A^{\rm dir}_{CP}(\Xi_{c}^{+}\to \Xi^{0}K^+)=0,\\
  &  A^{\rm dir}_{CP}(\Xi_{c}^{0}\to \Sigma^{-}\pi^+) +  A^{\rm dir}_{CP}(\Xi_{c}^{0}\to \Xi^{-}K^+)=0,\\
  &  A^{\rm dir}_{CP}(\Xi_{c}^{0}\to n\overline K^0) +  A^{\rm dir}_{CP}(\Xi_{c}^{0}\to \Xi^{0} K^0)=0,\\
  &  A^{\rm dir}_{CP}(\Xi_{c}^{0}\to \Sigma^{+}\pi^-) +  A^{\rm dir}_{CP}(\Xi_{c}^{0}\to pK^-)=0.
\end{align}
Under the complete interchange of $d \leftrightarrow s$, the doubly charmed baryons are interchanged as
\begin{align}
\Xi^{++}_{cc}\leftrightarrow \Xi^{++}_{cc},\quad \Xi^{+}_{cc}\leftrightarrow \Omega^{+}_{cc}.
\end{align}
The sum rules for $CP$ asymmetries in doubly charmed baryon decays into one pseudoscalar meson and one charmed triplet baryon are
\begin{align}
 & A^{\rm dir}_{CP}(\Xi_{cc}^{++}\to \Lambda^+_c\pi^+) +  A^{\rm dir}_{CP}(\Xi_{cc}^{++}\to \Xi^+_cK^+)=0,\\
  &  A^{\rm dir}_{CP}(\Xi_{cc}^{+}\to \Xi^+_cK^0) +  A^{\rm dir}_{CP}(\Omega_{cc}^{+}\to \Lambda^+_c\overline K^0)=0,\\
  &  A^{\rm dir}_{CP}(\Xi_{cc}^{+}\to \Xi^0_cK^+) +  A^{\rm dir}_{CP}(\Omega_{cc}^{+}\to \Xi^0_c\pi^+)=0.
\end{align}
Under the complete interchange of $d \leftrightarrow s$, the charmed sextet baryons are interchanged as
\begin{align}
\Sigma_c^+\leftrightarrow \Xi^{*+}_{c},\quad \Sigma^{++}_{c}\leftrightarrow \Sigma^{++}_{c},\quad \Xi^{*0}_{c}\leftrightarrow \Xi^{*0}_{c}, \quad \Sigma^{0}_{c}\leftrightarrow \Omega^{0}_{c}.
\end{align}
The sum rules for $CP$ asymmetries in doubly charmed baryon decays into one pseudoscalar meson and one charmed sextet baryon are
\begin{align}
 & A^{\rm dir}_{CP}(\Xi_{cc}^{++}\to \Sigma^+_c\pi^+) +  A^{\rm dir}_{CP}(\Xi_{cc}^{++}\to \Xi^{*+}_cK^+)=0,\\
  &  A^{\rm dir}_{CP}(\Xi_{cc}^{+}\to \Sigma^{++}_c\pi^-) +  A^{\rm dir}_{CP}(\Omega_{cc}^{+}\to \Sigma^{++}_cK^-)=0,\\
  &  A^{\rm dir}_{CP}(\Xi_{cc}^{+}\to \Sigma^0_c\pi^+) +  A^{\rm dir}_{CP}(\Omega_{cc}^{+}\to \Omega^0_cK^+)=0,\\
  &  A^{\rm dir}_{CP}(\Xi_{cc}^{+}\to \Xi^{*+}_cK^0) +  A^{\rm dir}_{CP}(\Omega_{cc}^{+}\to \Sigma^{+}_c \overline K^0)=0,\\
  &  A^{\rm dir}_{CP}(\Xi_{cc}^{+}\to \Xi^{*0}_cK^+) +  A^{\rm dir}_{CP}(\Omega_{cc}^{+}\to \Xi^{*0}_c\pi^+)=0.
\end{align}
With the interchange rules mentioned above,
the $CP$ violation sum rules in doubly charmed baryon decays into one charmed meson and one octet baryon are
\begin{align}
 & A^{\rm dir}_{CP}(\Xi_{cc}^{++}\to \Sigma^+D^+_s) +  A^{\rm dir}_{CP}(\Xi_{cc}^{++}\to pD^+)=0,\\
  &  A^{\rm dir}_{CP}(\Xi_{cc}^{+}\to pD^0) +  A^{\rm dir}_{CP}(\Omega_{cc}^{+}\to \Sigma^{+}D^0)=0,\\
  &  A^{\rm dir}_{CP}(\Xi_{cc}^{+}\to nD^+) +  A^{\rm dir}_{CP}(\Omega_{cc}^{+}\to \Xi^{0}D^+_s)=0.
\end{align}
The $CP$ violation sum rules in doubly charmed baryon decays into one charmed meson and one decuplet baryon are
\begin{align}
 & A^{\rm dir}_{CP}(\Xi_{cc}^{++}\to \Delta^+D^+) +  A^{\rm dir}_{CP}(\Xi_{cc}^{++}\to \Sigma^{*+}D^+_s)=0,\\
  &  A^{\rm dir}_{CP}(\Xi_{cc}^{+}\to\Delta^{+}D^0) +  A^{\rm dir}_{CP}(\Omega_{cc}^{+}\to \Sigma^{*+}D^0)=0,\\
  &  A^{\rm dir}_{CP}(\Xi_{cc}^{+}\to \Delta^{0}D^+) +  A^{\rm dir}_{CP}(\Omega_{cc}^{+}\to \Xi^{*0}D^+_s)=0,\\
  &  A^{\rm dir}_{CP}(\Xi_{cc}^{+}\to \Sigma^{*0}D^+_s) +  A^{\rm dir}_{CP}(\Omega_{cc}^{+}\to \Sigma^{*0}D^+)=0.
\end{align}
For three-body decays, we only list the $CP$ violation sum rules in charmed baryon decays into one octet baryon and two pseudoscalar mesons as examples:
\begin{align}
 & A^{\rm dir}_{CP}(\Lambda_{c}^{+}\to pK^-K^+) +  A^{\rm dir}_{CP}(\Xi_{c}^{+}\to \Sigma^+\pi^-\pi^+)=0,\\
& A^{\rm dir}_{CP}(\Lambda_{c}^{+}\to p\pi^-\pi^+) +  A^{\rm dir}_{CP}(\Xi_{c}^{+}\to \Sigma^+K^-K^+)=0,\\
& A^{\rm dir}_{CP}(\Lambda_{c}^{+}\to \Sigma^+\pi^-K^+) +  A^{\rm dir}_{CP}(\Xi_{c}^{+}\to pK^-\pi^+)=0,\\
& A^{\rm dir}_{CP}(\Lambda_{c}^{+}\to \Sigma^-\pi^+K^+) +  A^{\rm dir}_{CP}(\Xi_{c}^{+}\to \Xi^-K^+\pi^+)=0,\\
& A^{\rm dir}_{CP}(\Lambda_{c}^{+}\to nK^+\overline K^0) +  A^{\rm dir}_{CP}(\Xi_{c}^{+}\to \Xi^0\pi^+K^0)=0,\\
  &  A^{\rm dir}_{CP}(\Xi_{c}^{0}\to \Sigma^{+}K^-K^0) +  A^{\rm dir}_{CP}(\Xi_{c}^{0}\to p\pi^-\overline K^0)=0,\\
  &  A^{\rm dir}_{CP}(\Xi_{c}^{0}\to \Sigma^-K^+\overline K^0) +  A^{\rm dir}_{CP}(\Xi_{c}^{0}\to \Xi^{-}\pi^+ K^0)=0,\\
  &  A^{\rm dir}_{CP}(\Xi_{c}^{0}\to \Xi^{0}\pi^-K^+) +  A^{\rm dir}_{CP}(\Xi_{c}^{0}\to nK^-\pi^+)=0.
\end{align}
The first three sum rules are the same as in  \cite{Grossman:2018ptn}.
In all above sum rules, the pseudoscalar mesons can be replaced by vector mesons by the following correspondence:
\begin{align}
\pi^+\rightarrow \rho^+,\quad \pi^-\rightarrow \rho^-, \quad K^+\rightarrow K^{*+},\quad K^-\rightarrow K^{*-}, \quad K^0\rightarrow K^{*0},  \quad \overline K^0\rightarrow \overline K^{*0}.
\end{align}

The $CP$ violation sum rules are derived in the $U$-spin limit. Considering the $U$-spin breaking, the $CP$ violation sum rules are no longer valid, as pointed out in \cite{Grossman:2018ptn}.
Since the $U$-spin breaking is sizable in the charm sector, the $CP$ violation sum rules might not be reliable. But they indicate that the $CP$ asymmetries in some decay modes have opposite sign and then can be used to find some promising observables in experiments.
In charmed meson decays, Eq.~\eqref{b8} makes the two $CP$ asymmetries in observable $\Delta A_{CP} \equiv A_{CP}(D^0\to K^+K^-) - A_{CP}(D^0\to \pi^+\pi^-)$ constructive. Similarly, one can use the $CP$ violation sum rules in charmed baryon decays to construct some observables in which two $CP$ asymmetries are constructive.
Some observables are selected for experimental discretion:
\begin{align}
 \Delta A_{CP}^{\rm baryon,1}&= A_{CP}(\Lambda_{c}^{+}\to\Sigma^+K^{*0}) -  A_{CP}(\Xi_{c}^{+}\to p\overline K^{*0}),\\
 \Delta A_{CP}^{\rm baryon,2}&=A_{CP}(\Xi_{c}^{0}\to \Sigma^{+}\pi^-) -  A_{CP}(\Xi_{c}^{0}\to pK^-),\\
 \Delta A_{CP}^{\rm baryon,3}&=A_{CP}(\Lambda_{c}^{+}\to \Delta^{++}\pi^-) -  A_{CP}(\Xi_{c}^{+}\to \Delta^{++} K^-),\\
\Delta A_{CP}^{\rm baryon,4}&=A_{CP}(\Lambda^+_c\to \Sigma^+\pi^-K^+) -A_{CP}(\Xi^+_c\to pK^-\pi^+).
\end{align}

If the contributions proportional to $\lambda_b$ are neglected,
the decay amplitudes of the two channels connected by the interchange of $d \leftrightarrow s$ are the same (except for a minus sign) in the $SU(3)_F$ limit [see Eqs.~\eqref{x1} and ~\eqref{x2}].  One can use this relation to predict the branching fractions. As an example, we estimate the branching fraction of $\Xi^+_c\to pK^-\pi^+$.
The integration over the phase space of the three-body decay  $\mathcal{B}_c\to \mathcal{B}M_1M_2$ relies on the equation of \cite{Tanabashi:2018oca}
\begin{align}
 \Gamma(\mathcal{B}_c\to \mathcal{B}M_1M_2)=\int_{m_{12}^2}\int_{m_{23}^2}\frac{|\mathcal{A}(\mathcal{B}_c\to \mathcal{B}M_1M_2)|^2}{32m^3_{\mathcal{B}_c}}\mathrm{{d}}m^2_{12}\mathrm{{d}}m^2_{23},
\end{align}
where $m_{12}^2=(p_{M_1} + p_{M_2})^2$ and  $m_{23}^2=(p_{M_2} + p_{\mathcal{B}})^2$.
With the experimental data given in \cite{Tanabashi:2018oca},
\begin{align}
 \mathcal{B}r(\Lambda^+_c\to \Sigma^+\pi^-K^+) = (2.1\pm 0.6)\times 10^{-3},
\end{align}
and the relation
\begin{align}
|\mathcal{A}(\Xi^+_c\to pK^-\pi^+)| \simeq |\mathcal{A}(\Lambda^+_c\to \Sigma^+\pi^-K^+)|,
\end{align}
the branching fraction of $\Xi^+_c\to pK^-\pi^+$ decay is predicted to be
\begin{align}\label{b9}
 \mathcal{B}r(\Xi^+_c\to pK^-\pi^+) = (1.7\pm 0.5)\%.
\end{align}
One can find the branching fraction $\mathcal{B}r(\Xi^+_c\to pK^-\pi^+)$ is larger than $\mathcal{B}r(\Lambda^+_c\to \Sigma^+\pi^-K^+)$ because of the larger phase space and the longer lifetime of $\Xi^+_c$. But it is still smaller than the predictions given in \cite{Jiang:2018iqa,Geng:2018upx}.
In the above estimation, only the decay amplitude is obtained by the $U$-spin symmetry. The phase space is calculated without approximation. It is plausible since the global fit in Refs.~\cite{Hsiao:2019yur,Geng:2018rse,Geng:2018upx,Geng:2017esc,Geng:2017mxn,Geng:2018plk,Geng:2018bow}
give the reasonable estimations for branching fractions of charmed baryon decays.
The uncertainty in Eq.~\eqref{b9} is dominated by the branching fraction of $\Lambda^+_c\to \Sigma^+\pi^-K^+$ decay and does not include the $U$-spin breaking effects. It is not available to estimate the $U$-spin breaking effects at the current
stage since the understanding of the dynamics of charmed baryon decay is still a challenge. Some discussions of the uncertainty induced by $U$-spin breaking can be found in \cite{Jiang:2018iqa}.

With the method introduced in \cite{Jiang:2018iqa}, and the LHCb data \cite{Aaij:2014esa}
\begin{align}
\frac{f_{\Xi_b}}{f_{\Lambda_b}}\cdot\frac{\mathcal{B}r(\Xi^0_b\to \Xi^+_c\pi^-)}{\mathcal{B}r(\Lambda^0_b\to \Lambda^+_c\pi^-)}\cdot \frac{\mathcal{B}r(\Xi^+_c\to pK^-\pi^+)}{\mathcal{B}r(\Lambda^+_c\to pK^-\pi^+)} = (1.88\pm 0.04\pm 0.03 )\times 10^{-2},
\end{align}
the fragmentation-fraction ratio $f_{\Xi_b}/f_{\Lambda_b}$ is determined to be
\begin{align}
f_{\Xi_b}/f_{\Lambda_b}=0.065\pm 0.020.
\end{align}
Recent measurement confirmed this result \cite{Aaij:2019ezy}.
Our result is consistent with the one obtained via $\Lambda^0_b\to J /\psi \Lambda^0$ \cite{Voloshin:2015xxa}, $f_{\Xi_b}/f_{\Lambda_b}=0.11\pm 0.03$, the one via $\Xi^-_b\to J /\psi  \Xi^-$ \cite{Hsiao:2015txa}, $f_{\Xi_b}/f_{\Lambda_b}=0.108\pm 0.034$, and the one via the diquark model for $\Xi^-_b\to \Lambda^0_b\pi^-$ \cite{Cheng:2015ckx} using the LHCb data \cite{Aaij:2015yoy}, $f_{\Xi_b}/f_{\Lambda_b}=0.08\pm 0.03$. The detailed comparison for different methods of estimating $f_{\Xi_b}/f_{\Lambda_b}$ can be found in \cite{Jiang:2018iqa}.

\section{Summary}\label{co}
In summary, we find that if two singly Cabibbo-suppressed decay modes of charmed hadrons are connected by a complete interchange of $d$ and $s$ quarks, the sum of their direct $CP$ asymmetries is zero in the flavor $SU(3)$ limit. According to this conclusion,
many $CP$ violation sum rules  can be found in the doubly and singly charmed baryon decays.
Some of them could help to find better observables in experiments.
As byproducts, the branching fraction $\mathcal{B}r(\Xi^+_c\to pK^-\pi^+)$ is predicted to be  $(1.7\pm 0.5)\%$ in the $U$-spin limit, and the fragmentation-fraction ratio is determined as $f_{\Xi_b}/f_{\Lambda_b}=0.065\pm 0.020$.

\acknowledgments

We are grateful to Fu-Sheng Yu for useful discussions. This work was supported in part by  the National Natural Science Foundation of China under
Grants no. U1732101 and the Fundamental Research Funds for the
Central Universities under Grant no. lzujbky-2018-it33.

\begin{appendix}

\section{$SU(3)$ decomposition of operator $O^{ij}_k$}\label{dec}
All components of the $SU(3)$ decomposition in Eq.~\eqref{hd} are listed in the following.\\
$\overline 3$ presentation:
\begin{align}\label{3t}
 O(\overline 3)^1 & = (\bar u u)(\bar u c) + (\bar ud)(\bar d c) + (\bar us)(\bar s c),\quad
  O(\overline3)^2 = (\bar d u)(\bar u c) + (\bar dd)(\bar d c) + (\bar ds)(\bar s c),\nonumber \\
 O(\overline3)^3 & = (\bar s u)(\bar u c) + (\bar sd)(\bar d c) + (\bar ss)(\bar s c).
\end{align}
$\overline 3^\prime$ presentation:
\begin{align}\label{3p}
  O(\overline 3^\prime)^1 & = (\bar u u)(\bar u c) + (\bar dd)(\bar u c) + (\bar ss)(\bar u c),\quad
 O(\overline3^\prime)^2 = (\bar u u)(\bar d c) + (\bar dd)(\bar d c) + (\bar ss)(\bar d c),\nonumber \\
  O(\overline3^\prime)^3 & = (\bar u u)(\bar s c) + (\bar dd)(\bar s c) + (\bar ss)(\bar s c).
\end{align}
$6$ presentation:
\begin{align}
    O( 6)_{11} & = \frac{1}{2} [(\bar d u)(\bar s c) - (\bar su)(\bar d c)],\quad O( 6)_{22} = \frac{1}{2} [(\bar s d)(\bar u c) - (\bar ud)(\bar s c)],\nonumber \\
   O( 6)_{33} &= \frac{1}{2} [(\bar u s)(\bar d c) - (\bar ds)(\bar u c)],\nonumber \\
    O(6)_{12} & = \frac{1}{4} [(\bar s u)(\bar u c) - (\bar uu)(\bar s c) + (\bar dd)(\bar s c) - (\bar sd)(\bar d c)],\nonumber \\
    O( 6)_{23}& = \frac{1}{4} [(\bar u d)(\bar d c) - (\bar dd)(\bar u c) + (\bar ss)(\bar u c) -(\bar us)(\bar s c)],\nonumber \\
   O( 6)_{31} & = \frac{1}{4} [(\bar d s)(\bar s c) - (\bar ss)(\bar d c) + (\bar uu)(\bar d c) -(\bar du)(\bar u c)].
\end{align}
$ {\overline{15}}$ presentation:
\begin{align}\label{15}
     O(\overline{15})_{1}^{11} & = \frac{1}{2}(\bar u u)(\bar u c)-\frac{1}{4} [(\bar us)(\bar s c) + (\bar ud)(\bar dc)+(\bar dd)(\bar u c) + (\bar ss)(\bar uc)],\nonumber \\
     O(\overline{15})_{2}^{22} & = \frac{1}{2}(\bar dd)(\bar d c)-\frac{1}{4} [(\bar du)(\bar u c) + (\bar ds)(\bar sc)+(\bar uu)(\bar d c) + (\bar ss)(\bar dc)],\nonumber \\
     O(\overline{15})_{3}^{33} & = \frac{1}{2}(\bar ss)(\bar s c)-\frac{1}{4} [(\bar su)(\bar u c) + (\bar sd)(\bar dc)+(\bar uu)(\bar s c) + (\bar dd)(\bar sc)],\nonumber \\
    O(\overline{15})_{1}^{23} & = \frac{1}{2} [(\bar d u)(\bar s c) + (\bar su)(\bar d c)],\quad         O(\overline{15})_{2}^{13} = \frac{1}{2} [(\bar s d)(\bar u c) + (\bar ud)(\bar s c)],  \nonumber \\
    O(\overline{15})_{3}^{12} & = \frac{1}{2} [(\bar u s)(\bar d c) + (\bar ds)(\bar u c)],\nonumber \\
     O(\overline{15})_{2}^{11} & = (\bar u d)(\bar u c), \quad  O(\overline{15})_{3}^{11}  = (\bar u s)(\bar u c), \quad  O(\overline{15})_{1}^{22} = (\bar du)(\bar d c),\nonumber \\
     O(\overline{15})_{3}^{22} & = (\bar d s)(\bar d c), \quad O(\overline{15})_{1}^{33} = (\bar su)(\bar s c), \quad
      O(\overline{15})_{2}^{33}  = (\bar s d)(\bar s c),\nonumber \\
       O(\overline{15})_{1}^{21} & = \frac{3}{8}[(\bar u u)(\bar d c)+(\bar d u)(\bar u c)]-\frac{1}{4}(\bar d d)(\bar d c) - \frac{1}{8}[(\bar d s)(\bar s c)+(\bar ss)(\bar d c)],\nonumber \\
      O(\overline{15})_{2}^{12} & = \frac{3}{8}[(\bar u d)(\bar d c)+(\bar d d)(\bar u c)]-\frac{1}{4}(\bar u u)(\bar u c) - \frac{1}{8}[(\bar u s)(\bar s c)+(\bar ss)(\bar u c)],\nonumber \\
     O(\overline{15})_{1}^{31} & = \frac{3}{8}[(\bar u u)(\bar s c)+(\bar su)(\bar u c)]-\frac{1}{4}(\bar s s)(\bar s c) - \frac{1}{8}[(\bar s d)(\bar d c)+(\bar dd)(\bar s c)],\nonumber \\
     O(\overline{15})_{3}^{13} & = \frac{3}{8}[(\bar u s)(\bar s c)+(\bar ss)(\bar u c)]-\frac{1}{4}(\bar u u)(\bar u c) - \frac{1}{8}[(\bar u d)(\bar d c)+(\bar dd)(\bar u c)],\nonumber \\
       O(\overline{15})_{2}^{32} & = \frac{3}{8}[(\bar d d)(\bar s c)+(\bar sd)(\bar d c)]-\frac{1}{4}(\bar s s)(\bar s c) - \frac{1}{8}[(\bar s u)(\bar u c)+(\bar uu)(\bar s c)],\nonumber \\
      O(\overline{15})_{3}^{23} & = \frac{3}{8}[(\bar ds)(\bar s c)+(\bar s s)(\bar d c)]-\frac{1}{4}(\bar d d)(\bar dc) - \frac{1}{8}[(\bar d u)(\bar u c)+(\bar uu)(\bar d c)].
\end{align}
The above results are consistent with Ref.~\cite{Grossman:2012ry}.
The operators $O(6)_{ij}$ are symmetric in the interchange of their two indices and can be
written as $O(6)^{ij}_k$ by contracting
with the totally antisymmetric Levi-Civita tensor, $O( 6)^{ij}_k=\epsilon^{ijl}O(6)_{lk}$.
There are 18 operators in the irreducible representation $\overline{15}$, but only 15 of them are independent because of the following equations:
\begin{align}
  O(\overline{15})_1^{11} &= -[O(\overline{15})_2^{12} +O(\overline{15})_3^{13}], \quad O(\overline{15})_2^{22} = -[O(\overline{15})_1^{21} +O(\overline{15})_3^{23}],\nonumber \\
  O(\overline{15})_3^{33}&= -[O(\overline{15})_1^{31} +O(\overline{15})_2^{32}].
\end{align}

\end{appendix}

\end{document}